\documentclass[conference]{IEEEtran}
\ifCLASSINFOpdf
\else
\fi
\usepackage{graphicx}
\usepackage{cite}
\usepackage{algorithm}
\usepackage[noend]{algpseudocode}
\usepackage{algorithmicx}
\usepackage[draft]{hyperref}
\usepackage{subcaption}

\usepackage{amsmath}
\usepackage{amssymb}
\begin{document}
\bstctlcite{IEEEexample:BSTcontrol}
%
\title{Semantic Routing for Enhanced Performance of LLM-Assisted Intent-Based 5G Core Network Management and Orchestration
}
%
%
%

\author{Dimitrios~Michael~Manias, Ali~Chouman, and~Abdallah~Shami \\ The Department of Electrical and Computer Engineering, Western University\\ \{dmanias3, achouman, Abdallah.Shami\}@uwo.ca}

\maketitle

\begin{abstract}
Large language models (LLMs) are rapidly emerging in Artificial Intelligence (AI) applications, especially in the fields of natural language processing and generative AI. Not limited to text generation applications, these models inherently possess the opportunity to leverage prompt engineering, where the inputs of such models can be appropriately structured to articulate a model's purpose explicitly. A prominent example of this is intent-based networking, an emerging approach for automating and maintaining network operations and management. This paper presents semantic routing to achieve enhanced performance in LLM-assisted intent-based management and orchestration of 5G core networks. This work establishes an end-to-end intent extraction framework and presents a diverse dataset of sample user intents accompanied by a thorough analysis of the effects of encoders and quantization on overall system performance. The results show that using a semantic router improves the accuracy and efficiency of the LLM deployment compared to stand-alone LLMs with prompting architectures.
\end{abstract}


%
\IEEEpeerreviewmaketitle

\section{Introduction}
%
%
%
%
\IEEEPARstart{I}{ntent}-based networking research has seen various new Large Language Model (LLM) architectures and spurred the need for fine-tuning models, especially in the 5G context \cite{mcnamara2023nlp}. These applications often depend on the use of closed-source LLMs. While traditionally achieving better performance than open-source models, closed-source LLMs present certain risks and concerns related to their composition and functional behaviour. Recent advancements in the state-of-the-art position open-source LLMs, such as LLaMa-2 or Mistral 7B, to perform comparably to Open AI's closed-source GPT models. Additionally, the concept of model tractability is highly advantageous for open-source solutions \cite{laskar2023building}. In addition, singling out the best LLM, in terms of performance, is challenging due to the nature of development in closed-source and open-source. Closed-source LLMs are updated by continually retraining on newer data, and open-source LLMs are released in response to match their performance \cite{chen2023chatgpt}. There is a lack of transparency in identifying model or process changes in closed-source LLM development, limiting their usefulness in privacy-sensitive applications. Conversely, open-source LLMs are not limited to being LLaMa-based, but can also be trained from scratch, such as Falcon, Phi, MPT, and other generic LLMs \cite{zafar2023building}. Overall, the limitation of generalization across open-source and closed-source LLMs, respective sample sizes, access to training data, and inherent bias, present an imperative journey toward the democratization of LLM research, development, and incorporation.

Furthermore, there are privacy risks associated with closed-source LLMs only accessible via API, which can be bypassed by adversarial strategies that compromise or breach LLM privacy \cite{wu2024new}. Mitigation, for example, is possible for LLMs such as GPT-4 by programming the restriction of executing external instructions without user confirmation. GPT-4 also sends HTTP requests to target links generated by the LLM to retrieve content, which can lead to private data leakage. Implementing a safe URL check procedure can also address this data leakage issue for GPT-4 and other LLMs. Moreover, there are additional practices that closed-source LLMs should adopt concerning their own evaluation. In practice, LLM evaluations should be reproducible, and the performance and output of such models should be interpreted with caution. Also, the indirect data leaking of proprietary LLMs should be monitored and reported to spur reviewers to demand a transparent and objective evaluation of LLM research \cite{balloccu2024leak}.

The work presented in this paper promotes the use of open-source models for enhanced LLM-assisted Intent-based Network Management and Orchestration (MANO) in the 5G Core. Specifically, using the Mistral 7B model alongside a semantic router provides time and accuracy efficiencies compared to existing LLM deployments. Also, an evaluation of LLM quantization is explored to address a major criticism of LLMs: the resources required to train, optimize, and host them. Finally, a diverse dataset containing intents with expectations for the 5G network is generated to improve the performance of the proposed solution and catalyze research in the field.

\section{Related Work}
Intent-based networking bridges the gap between end-user service requests and network operations. Leivadeas and Falkner \cite{leivadeas2022survey} have extensively surveyed the field and have provided a taxonomy of the various methods intent-based networking can be achieved, including template-based, NLP-based, custom language-based, API-base, and Grammar/Keyword-based. Of these identified methods, the NLP-based methods provide the highest level of flexibility, allowing intents to be expressed in conventional human language. Traditionally, models such as BERT have been used to address NLP tasks, including name entity recognition, relation extraction, and keyword extraction, which have been leveraged in NLP-based intent-based networking. McNamara \textit{et al.} present an NLP-based interface for intent-based networking in private 5G networks using the lightweight Adaptive Policy Execution (APEX) engine \cite{mcnamara2023nlp}. The authors identify the use of LLMs to simplify the intent mapping workflow. Dzeparoska \textit{et al.} leverage LLMs for an intent fulfilment and assurance framework and define intent drift through KPI monitoring \cite{dzeparoska2024intent}. Wang \textit{et al.} directly apply LLMs to network packet data to develop a knowledge space that can be used for traffic perception, network modelling, and management insights \cite{wang2023network}. The work presented differs from the state-of-the-art as it considers explicitly the intricate nature of the 5G core network MANO and aligns with the most recent 3GPP standards.
Our previous work presented a standalone LLM agent based on prompting to perform intent extraction \cite{manias2024towards}. In this work, it was evident that the LLM could correctly identify the presence of various intents in a single request; however, it faced challenges in distinguishing between intents of a similar nature. A natural solution would be to adjust the prompt to contain additional information and examples of each intent, hoping to provide a more accurate extraction. Unfortunately, given the token limits of LLM inputs, this method is not scalable and would be unsustainable as intent types and complexities increase. Furthermore, after a period of operation, the LLM agent in question exhibited signs of hallucination, with incorrect, nonsensical outputs being returned. 

To this end, the work presented in this paper leverages a semantic router to enhance the reliability of the intent extraction process through deterministic routing. The contributions of this work can be summarized as follows:
\begin{itemize}
\item The introduction of an end-to-end intent extraction framework that leverages the semantic router to improve the performance and reliability of LLM-assisted intent-based networking in 5G Core MANO.
\item The creation of a linguistically diverse dataset containing examples of the various 3GPP-defined 5G core network intents.
\item An extensive analysis of the performance of the semantic router and the effect of linguistic diversity, encoding, and quantization on the framework’s performance.
\end{itemize}

\section{Background}

\textbf{Large Language Models:}
Since their inception, LLMs have taken the world by storm through their enhanced linguistic capabilities compared to their predecessors. LLMs, such as Open AI’s ChatGPT, were significant disruptors in the tech space as they removed a barrier to accessing AI for all. Since their initial introduction, the industrial and academic communities have been actively working to harness their potential through innovative startups and lucrative projects. Despite their incredible capabilities, caution must be taken to fully understand their shortcomings and limitations, ranging from ethical considerations to hallucinations, where an LLM generates nonsensical or blatantly incorrect outputs. Demystifying the aura around LLMs and genuinely understanding the extent of their capabilities is a critical step towards effectively integrating them into lucrative use cases and harnessing their incredible potential.

\textbf{Semantic Router:}
The semantic router is a method to introduce stability and reliability into an LLM deployment through deterministic decision-making. Essentially, the semantic router is a layer introduced into the LLM deployment that, as its name suggests, uses semantic meaning to route an input to the desired output. The semantic router's advantage over a standalone LLM is its ability to define explicit routes without performance deterioration due to LLM hallucinations, provide near-instant responses, and develop an end-to-end function calling pipeline. Furthermore, the semantic router can also be used as a system guardrail since each invoked route's resulting action can be tailored to ensure ethical compliance. Finally, the scalability aspect of the semantic router to scale to hundreds of thousands of routes and its possible integration with vectorized databases to host examples of each route work at overcoming the input token limitation faced by standalone LLMs.

\textbf{Encoders:}
The encoder of the semantic router is used to create text embeddings that establish relationships between words using distance-based metrics. Many different types of embedding tools exist, both open and closed source. In this work, two encoders are tested, OpenAI's \textit{text-embedding-ada-002} and Hugging Face's \textit{all-MiniLM-L6-v2}. The encoder's choice can affect the semantic router's performance; for example, the \textit{all-MiniLM-L6-v2} is a lightweight encoder intended for small paragraphs and truncates anything beyond 256 words, whereas the \textit{text-embedding-ada-002} can accept larger inputs.
 
\textbf{Quantization:}
LLM quantization is one of the emerging methods of democratization in the field of open-source LLMs. Traditionally, hosting, using, and fine-tuning these models has required excessive amounts of computational resources. Quantization effectively reduces the resource requirements by reducing the precision of the LLM’s weights. There are various methods of quantization, each with its pros and cons. In general, the more quantization is leveraged, the fewer resources are required to host and use the model; however, quantization also leads to a loss in the quality of the LLM in terms of performance. Exploring the quantization level suitable for a use case is important in determining LLM feasibility and applicability, especially in resource-constrained settings.

\textbf{Intent-Based Networking in the 5G Core:}
3GPP Technical Specification 28.312 (ETSI TS 128 312) \cite{intent_standard} defines a set of intent-driven management services for mobile networks. The latest version of this specification considers 5G Release 18. Based on this specification, the types of intents related to the 5G core network can be defined as follows: \textit{Deployment Intent}, \textit{Modification Intent}, \textit{Performance Assurance Intent}, \textit{Intent Report Request}, \textit{Intent Feasibility Check}, and \textit{Regular Notification Request}. Intent-based networking in the 5G core should leverage a framework that can ensure the end-to-end fulfilment of these intents, from identifying what intent is present to determining the appropriate action required to ensure the completion of the intent in an automated manner.

\section{Methodology}
\subsection{System Model}
\begin{figure}[!htbp]
\centerline{\includegraphics[width=1.0\columnwidth]{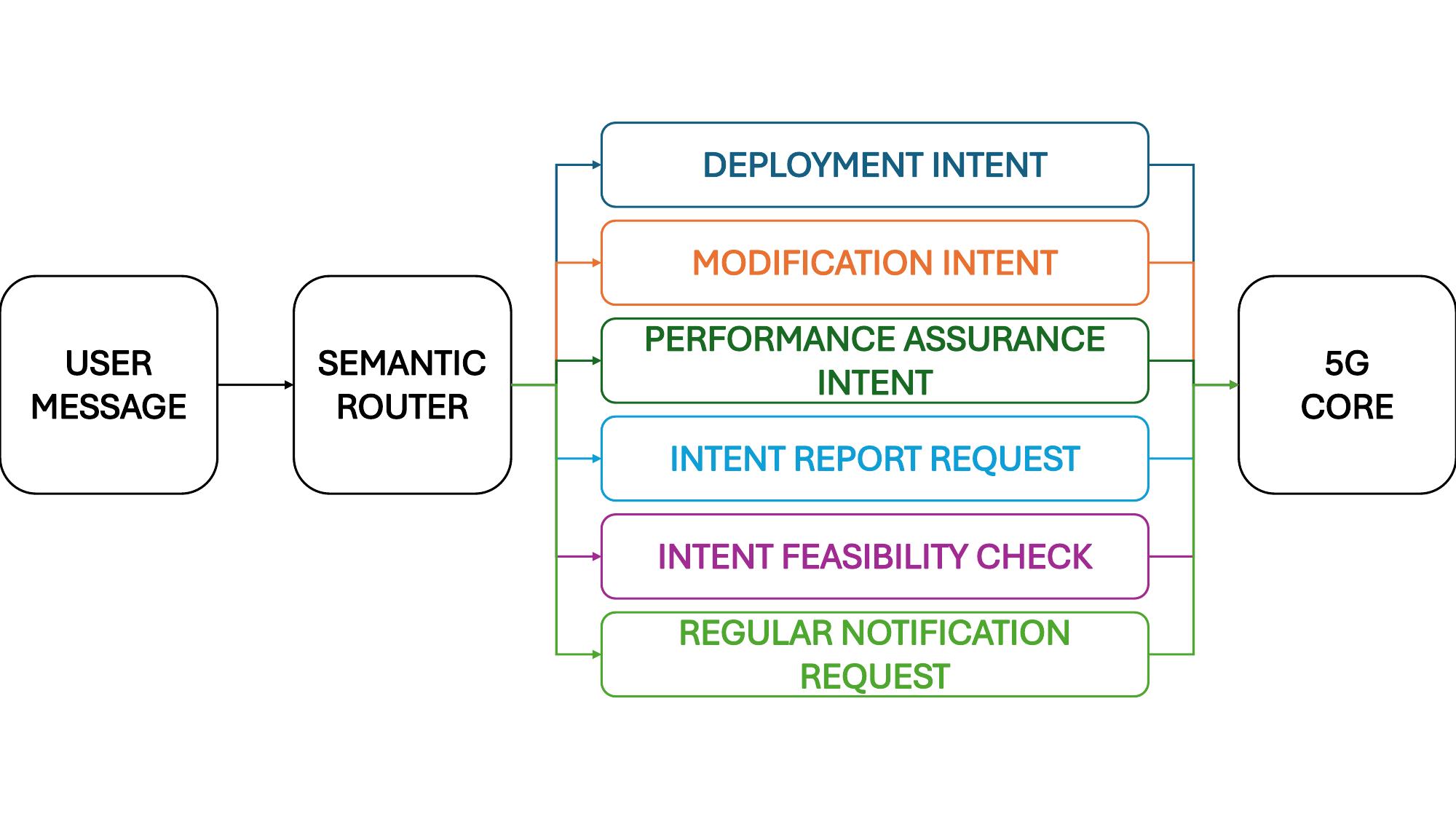}}
\caption{Semantic routing integration in 5G Core Intent-Based MANO Pipeline}
\label{sys_mod}
\end{figure}

The system model of the proposed framework is presented in Fig. \ref{sys_mod}. This system has three main entities: the user, the semantic router, and the 5G core network. This framework aims to provide end-to-end MANO capabilities through intent extraction. The first stage of this framework involves the user's message. This message starts the intent interpretation process and is passed to the semantic router. The objective of the semantic router is to interpret the user's intent and route the information through one of its preconfigured routes. For this work, six static routes have been configured (and by default, a ``None" route exists). Each of these routes is preconfigured to take a set of actions based on the requirements and nature of the intent. As such, handling each intent becomes independent, modular, and automated. Ultimately, each route will interact with the 5G Core and its management entities.

\subsection{Prompt Generation}
This section describes the various prompt-generating steps used throughout this work. This work dealt with three main types of prompts: seed, variability, and paraphrased. Pending this work's acceptance, the generated prompts will be made publicly available as a dataset on the OC2 Lab's GitHub page\footnote{\url{https://github.com/Western-OC2-Lab/}} in an effort to democratize research in this field. 

\subsubsection{Seed Prompts}
The seed prompts used in this work are a set of manually crafted prompts containing a single intent in accordance with the 3GPP/ETSI technical specifications. These prompts were created to add as much diversity as possible while including common conventions seen in the 5G core network space. Some examples of diversity present in the seed prompts include:
\begin{itemize}
\item Location: City, Region, Geographic Coordinates, Data Center
\item Entity Identifiers: Name, Short-hand notation, Instance \#, Alphanumeric, Hexadecimal, UUID-based
\end{itemize}
Given the author’s extensive understanding of 5G core network entities and MANO actions, careful consideration was taken to ensure the maximum amount of information diversity is present for each intent. An example of a seed prompt for the Modification Intent is shown in Fig. \ref{prompt_type}.
\subsubsection{Variability Prompts}
The variability prompts were generated by taking each seed prompt and leveraging an LLM to transform the prompt. In the case of the variability prompts, the objective is to introduce linguistic variability by adjusting the wording or phrasing of the seed prompt. For this work, OpenAI’s ChatGPT 3.5 was used with the following instruction to achieve this \textit{“I need to introduce linguistic variability to the following prompts. Adjust the wording and phrasing as required.”} After providing the instruction and the seed prompts, the output was a set of new prompts with linguistic variability introduced. An example of a variability prompt is seen in Fig. \ref{prompt_type}. When compared to the initial seed prompt, it can be seen that wording changes (\textit{i.e.,} “configuration parameters” $\rightarrow$ “parameter settings” and “enhance throughput” $\rightarrow$ “boost data transfer rates”) were made, but the overall sentence structure and semantic meaning remains the same.
\subsubsection{Paraphrased Prompts}
\begin{figure}[!htbp]
\centerline{\includegraphics[width=1.0\columnwidth]{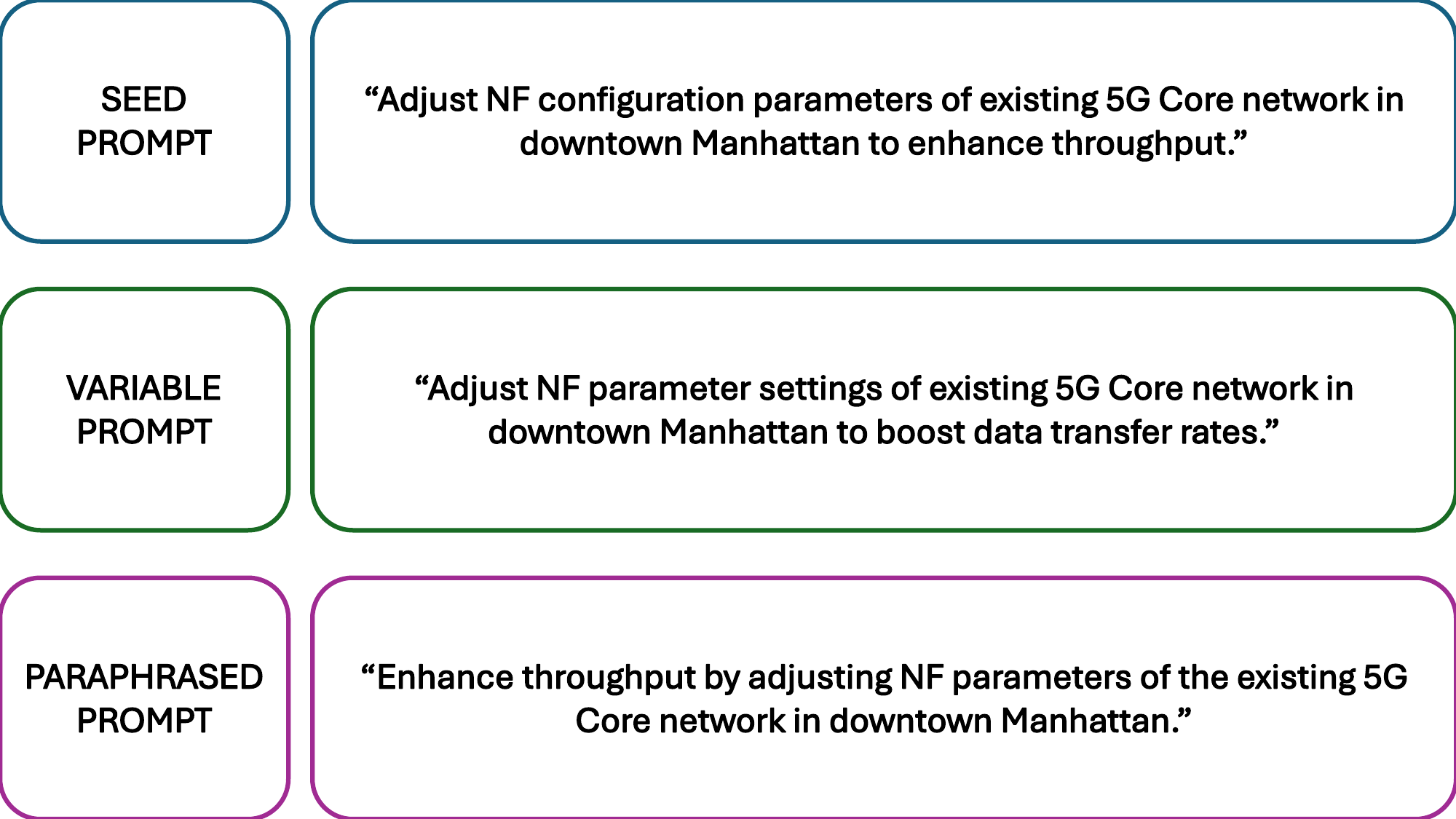}}
\caption{Exploring the various prompt transformation methods used.}
\label{prompt_type}
\end{figure}

In a similar fashion to the variability prompts, each of the paraphrase prompts was generated by leveraging LLMs to transform the seed prompts. In the case of paraphrased prompts, the objective is to change the sentence structure and wording of the prompt while keeping the same semantic meaning. GPT 3.5 was used with the following instruction to achieve this \textit{“I need to paraphrase the following prompts. Make sure to keep the same semantic meaning but change sentence structure and wording accordingly.”} Figure \ref{prompt_type} shows an example of the paraphrased seed prompt from before. In the initial seed prompt, the sentence is structured by first giving the action (adjust), then the entity location (downtown Manhattan), and then the motivation (enhance throughput). In the paraphrased prompt, it can be seen that the sentence is structured differently by first introducing the motivation, then describing the action, and then the entity location. This sentence structure change can lead to linguistic nuances and enhanced diversity through phrasing.

\subsection{Semantic Router Implementation}
For the purposes of this work, the Semantic Router, an open-source implementation developed by Aurelio AI, was used for the experiments \cite{semantic_router}. The process of creating the routing layer first requires defining a set of routes. For the initial stage of this work, static routes are defined; however, in future iterations, fully dynamic routes with direct function calling capabilities will be developed and integrated directly with the live end-to-end 5G testbed previously presented \cite{chouman2024modular}. The definition of a route requires a route name and a set of route utterances. The utterances provide the routing layers with examples of what should invoke the selected route and allow the framework to leverage the semantic vector space such that the semantic meaning of these routes can be generalized. The number of utterances can be scaled, something which is explored through the developed experiments. This work created six routes corresponding to the identified intents in the 3GPP/ETSI technical standards. Each of these routes has an initial base utterance corresponding to the example provided in the technical standard. The route names and their base utterances are presented in Table \ref{stand}. The semantic routing layer is defined by specifying an encoder to use, the set of routes, and the LLM to leverage. It should be noted that once defined, the routing layer can be used as a stand-alone entity.

\begin{table}[!htbp]
\caption{Route names and base utterances.}
\label{stand}
\begin{tabular}{|p{3.8cm}|p{4.2cm}|}
\hline
\textbf{Route Name}                   & \textbf{Base Utterance}                                                                                         \\ \hline
\textbf{Deployment Intent}            & ``Deploy a new network in {[}region{]} with the following specifications..."                                     \\ \hline
\textbf{Modification Intent}          &
``Modify the existing {[}network{]} to address the performance issues caused by high loading..."                 \\ \hline
\textbf{Performance Assurance Intent} & 
``Ensure that the deployed network can support a {[}QoS Level{]} application with the following requirements..." \\ \hline
\textbf{Intent Report Request}        & ``Summarize the results of the previous request."                                                                \\ \hline
\textbf{Intent Feasibility Check}     & ``Before proceeding, ensure that capacity exists in {[}region{]} to perform the required changes."               \\ \hline
\textbf{Regular Notification Request} & ``Notify me of the status of {[}network{]} every {[}frequency{]}."                                               \\ \hline
\end{tabular}
\end{table}

\subsection{Threshold Tuning}
Threshold tuning is a method of route layer optimization for the semantic router and works by providing a set of training utterances with their associated routes. The process works by determining the best thresholds for each route. It should be noted that the post-training threshold values are highly dependent on the type of encoder being used. By default, all routes begin with a threshold of 0.5.

\subsection{Experiment Data and Metrics}
Each experiment uses a dataset consisting of 30 seed prompts for each route, resulting in 180 total samples. 5-fold cross-validation is used in each experiment, and the presented results are averaged results across all folds. It should be noted that in the experiments where the number of utterances is increased, the utterances are removed from this data set and used as utterance data. This utterance data, in certain trials, is augmented using linguistic variability and paraphrasing methods, as previously discussed. Given the balanced nature of the dataset, only the accuracy results of each trial are presented. It should be noted that confusion matrices and an extensive set of derived metrics were calculated as part of the analysis; however, for brevity, only accuracy is shown.

\subsection{Experiments}
The following is a summary of the experiments conducted as part of this work. The base model used for these experiments, unless otherwise stated, is a 2-bit quantized version of the Mistral-7B-Instruct-v0.2 LLM \cite{quantized} and will be referred to as the Mistral 7B model throughout this section.
\subsubsection{Utterance Experiments}
This experiment explores the effect of varying the number of utterances passed to the semantic router on the pre and post-threshold training performance. This experiment uses the Mistral 7B model, the Hugging Face encoder, and utterances were varied to include 0, 15, 30, and 45 samples in addition to the initial utterance from the technical standards. It should be noted that in the case of 15 utterances, five seed prompts were used with their variability and paraphrased versions totalling 15. The same was done in the cases of 30 (10 seed prompts) and 45 (15 seed prompts).
\subsubsection{Vocabulary Diversity Experiments}
The objective of this experiment is to explore the effect of using variable language in addition to the seed prompts. For the purpose of these experiments, the Mistral 7B model with the Hugging Face encoder was used. The composition of the utterances was varied to explore cases where only seed prompts, a combination of seed and variable prompts, a combination of seed and paraphrased prompts, and a combination of all three prompts are used.
\subsubsection{Encoder Experiments}
This experiment's objective is to explore the encoder's effect on the system's performance. For these experiments, the Mistral 7B model was used along with either the Hugging Face or OpenAI encoders.
\subsubsection{Prompting vs. Routing Experiments}
The objective of this experiment is to see the impact of using a semantic router-based framework compared to a prompting-based framework for intent-based 5G core MANO. In this experiment, the prompting architecture explored in previous work is directly compared to the proposed framework, and each method's ability to correctly identify the presented intents is compared.
\subsubsection{Quantization Experiments}
The objective of this experiment is to explore the effect of LLM quantization on the semantic router framework's performance. In this experiment, various quantized versions of the Mistral 7B model \cite{quantized} were explored, including the following: Q2\_K: 2 bits 3.08 GB, Q4\_K\_S: 4 bits 4.14 GB, and Q6\_K: 6 bits 5.94 GB.

\section{Results and Analysis}

\subsection{Utterance Analysis}
\begin{figure}[!htbp]
\centerline{\includegraphics[width=0.75\columnwidth]{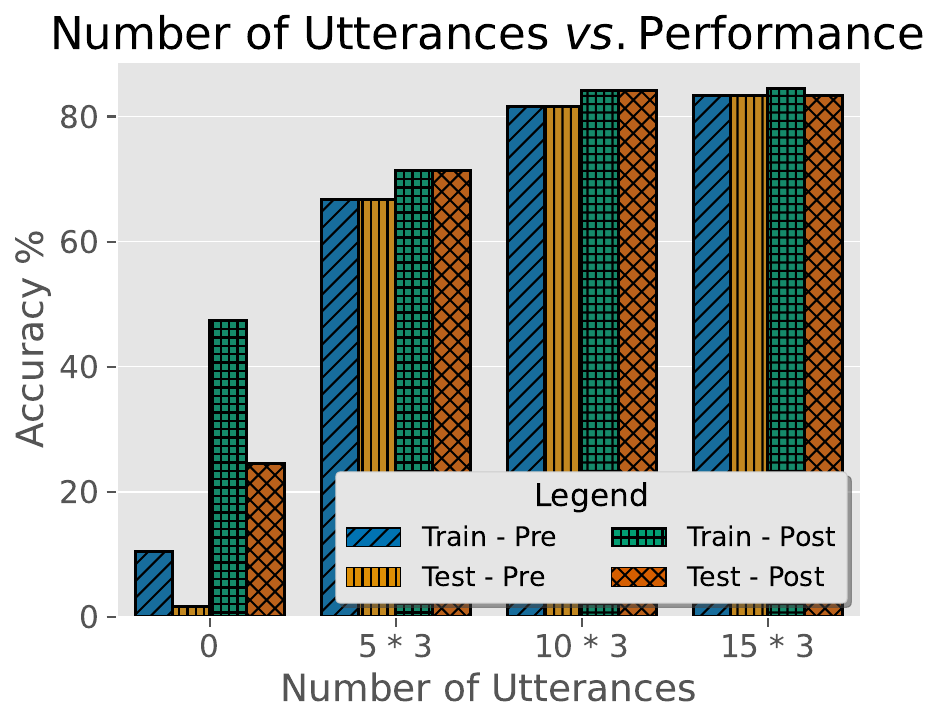}}
\caption{Semantic router with the Hugging Face encoder before and after route threshold training based on number of utterances provided.}
\label{utter}
\end{figure}

The results of the utterance experiments are presented in Fig \ref{utter}. These results display the semantic router's pre and post-threshold training performance on the train/test data. It should be noted that for the remainder of the results section, the notation $n * m$ denotes the number of $n$ seed prompts and $m$ versions (seed, variability, paraphrase) used in addition to the base descriptor prompt (from the standard) in each experiment. Looking at these results, it is evident that as the number of utterances increases, so too does the performance of the system. Intuitively, this makes sense as the more examples of the semantic meaning related to each type of intent are available, the greater the system's ability to generalize becomes. Another key insight is that fine-tuning the routing layer's threshold values improves the system's performance, as the post-training performance is greater than the pre-training performance on both the training and test sets. A final critical observation stemming from this experiment is that the increased use of utterances makes the performance more consistent across the train and test sets, meaning that the framework is able to generalize better to unseen samples. Evidence of this is seen as the gap between the train and test sets, when the number of utterances is increased, is reduced compared to the initial disparity seen when no additional utterances are included.

\subsection{Vocabulary Diversity Analysis}
\begin{figure}[!htbp]
\centerline{\includegraphics[width=0.75\columnwidth]{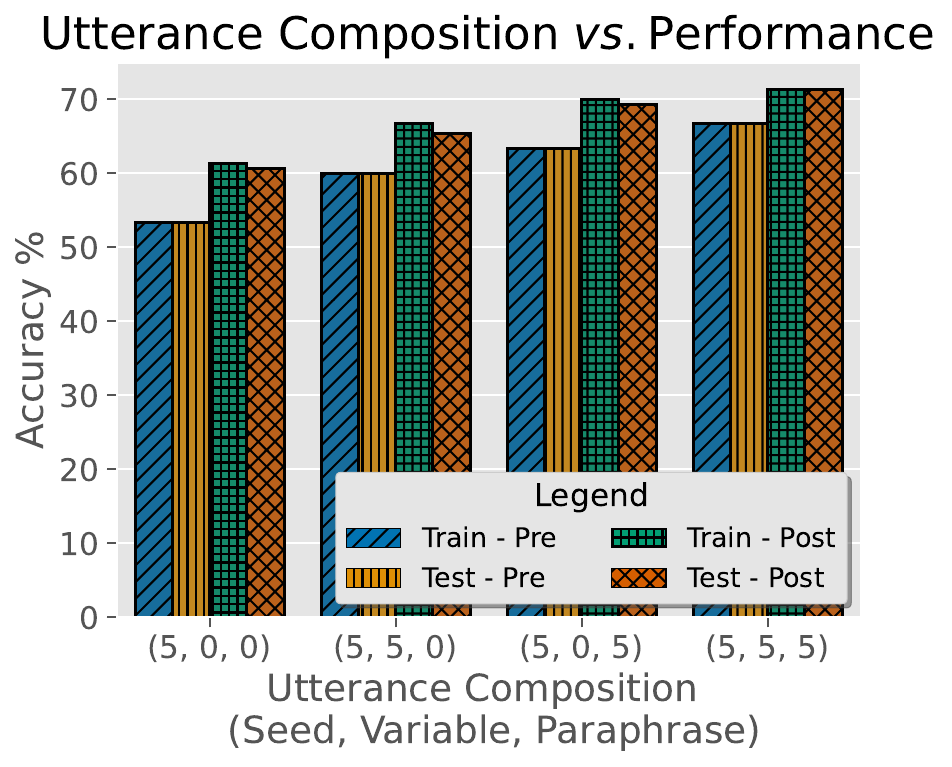}}
\caption{Semantic router with the Hugging Face encoder before and after route threshold optimization when varying the utterance vocabulary composition.}
\label{comp}
\end{figure}

The results of the vocabulary diversity experiments are presented in Fig. \ref{comp}. In these results, the notation $(a, b, c)$ denotes the number of $a$ seed prompts, $b$ variable prompts, and $c$ paraphrased prompts included as part of the semantic router's utterances. As seen in Fig. \ref{comp}, the linguistic diversity of the utterances plays a significant role in improving the system's performance despite conveying the same semantic meaning. Initially, when only five seed prompts are used, the maximum achievable performance is seen to be around 60\%. The inclusion of variable prompts or paraphrased prompts in addition to the seed prompts increases the performance by an additional 6\% and 8\%, respectively. The ideal case is seen when all three versions of the prompts are present, resulting in an overall increase of an additional 10\% in accuracy.

\subsection{Encoder Analysis}
\begin{figure}[!htbp]
\centerline{\includegraphics[width=0.75\columnwidth]{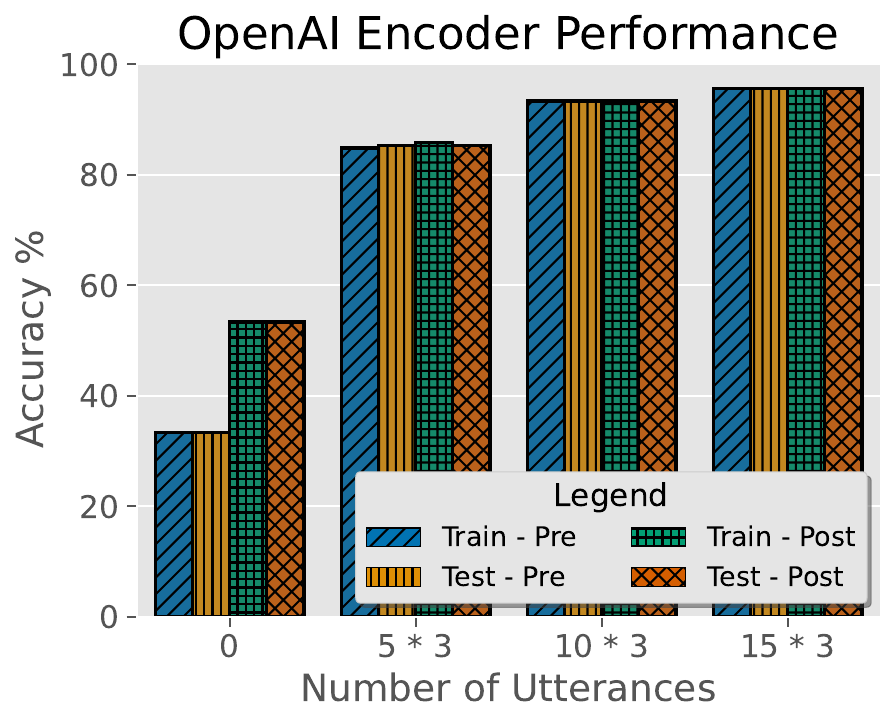}}
\caption{Semantic router with the OpenAI encoder before and after route threshold training based on number of utterances provided.}
\label{openai}
\end{figure}

The results when using the OpenAI encoder are presented in Fig. \ref{openai} and should be compared to Fig. \ref{utter}, which shows the same experiment using the Hugging Face encoder. Evidently, the OpenAI encoder outperforms the Hugging Face encoder in all trials. The OpenAI encoder performs significantly better even when no additional utterances other than the standard's base examples are provided. Additionally, beyond the 15 utterances, the threshold tuning ceases to affect the performance as there is no change between the threshold values and performance as a result of the training. These results present a critical trade-off between the two encoders that must be considered. The OpenAI encoder presents significant performance improvements over the Hugging Face encoder; however, it is a closed-source paid service as opposed to the open-source nature of the Hugging Face encoder. This has serious operational implications not just from a financial but also from an ethical perspective, as no insight into the inner workings of the OpenAI encoder nor its development process is available.

\subsection{Prompting vs. Routing Analysis}
\begin{figure}[!htbp]
\centerline{\includegraphics[width=0.7\columnwidth]{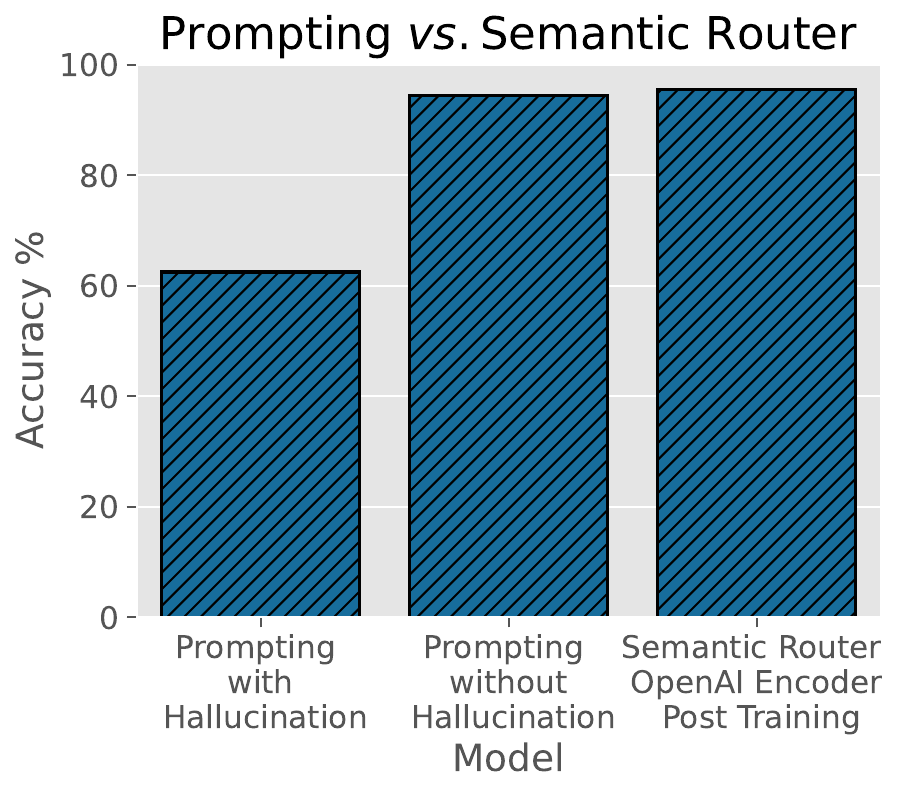}}
\caption{Comparison of semantic router and prompting.}
\label{prompting}
\end{figure}

The performance comparison between the semantic router framework and the previously proposed prompting architecture is presented in Fig. \ref{prompting}. These results show that the semantic router framework outperforms the prompting architecture purely from an accuracy perspective. It should be noted that there are two results shown for the prompting architecture, one with hallucination and the other without hallucination. LLM hallucination is a problem that plagues LLM deployments and is akin to catastrophic forgetting in conventional ML. Initially, the prompting architecture performed well; however, over time, its performance began degrading as unexpected outputs were returned. For example, despite being explicitly told the intent categories in the initial prompt, it began changing the category names. One example of this was the Performance Assurance Intent, which was referred to as ``Performance Intent" or ``Intent Assurance". In the presence of hallucination, the performance of the system is severely degraded. The semantic router is also better from an operational perspective as it takes a fraction of the time to return a result and is almost instantaneous. Empirically speaking, a time efficiency of 50x is observed when using the semantic router instead of the prompting architecture.

\subsection{Quantization Analysis}
The final set of results pertains to the quantization analysis. This experiment explored the use of models with various quantization levels to make them more feasible from a deployability perspective. The initial hypothesis, concurrent with the performance of other quantized models, is that the performance would be negatively impacted as greater levels of quantization were used. To our surprise, however, the results remained consistent across all levels of model quantization, and the smallest model performed just as well as the largest models tested. This presents a pivotal moment in making the case for the integration of LLMs in 5G core networks, as the reduction of model size without performance impacts in this setting addresses one of the main points of contention for LLM adoption in the field.
\section{Conclusion}

The work presented in this paper has made a compelling case for the use of the semantic router to increase the performance and reliability of LLM-based intent classification in 5G core networks. Through various experiments, a comprehensive set of insights into the effect of linguistic diversity, encoder type, and LLM quantization have been explored. Additionally, a direct performance comparison with a standalone prompting method has shown that the semantic router can perform in an expected and deterministic manner compared to the standalone LLM, which suffers from hallucinations over periods of extended use.
There exist several avenues for future work in this area. Firstly, regarding linguistic techniques to enhance seed prompts, this work considered variability and paraphrasing. A further extension of this, including methods such as back translation and tone shift, will be considered. Furthermore, a more comprehensive analysis of the performance of various combinations of LLMs with differing quantization levels and encoders must be explored. Another option is to use Retrieval Augmented Generation (RAG) for extensive collections of routes and route utterances. Moreover, as mentioned in the paper, converting the static routes into dynamic routes to perform live function calling will enable integration with a live 5G network. Finally, exploring the effect of multiple intents present per request and the best ways to handle the combination of multiple intents will be explored.

\bibliographystyle{IEEEtran}
\bibliography{sample}

\end{document}